\documentclass[manuscript]{emulateapj}

\usepackage{natbib}
\usepackage{xcolor}
\usepackage[]{graphics}
\usepackage{multirow}
\usepackage{rotating}  
\usepackage{lineno}

\shorttitle{Observations of M31 and M33 with the \textit{Fermi}-LAT}
\shortauthors{The \textit{Fermi}-LAT collaboration}

\newcommand{\fermi}{{\it Fermi}}

\begin{document}



\title{Observations of M31 and M33 with the \textit{Fermi} Large Area Telescope: \\ A galactic center excess in Andromeda?}

\author{
M.~Ackermann\altaffilmark{1}, 
M.~Ajello\altaffilmark{2}, 
A.~Albert\altaffilmark{3}, 
L.~Baldini\altaffilmark{4}, 
J.~Ballet\altaffilmark{5}, 
G.~Barbiellini\altaffilmark{6,7}, 
D.~Bastieri\altaffilmark{8,9}, 
R.~Bellazzini\altaffilmark{10}, 
E.~Bissaldi\altaffilmark{11}, 
E.~D.~Bloom\altaffilmark{12}
% ...
}

\altaffiltext{1}{Deutsches Elektronen Synchrotron DESY, D-15738 Zeuthen, Germany}
\altaffiltext{2}{Department of Physics and Astronomy, Clemson University, Kinard Lab of Physics, Clemson, SC 29634-0978, USA}
\altaffiltext{3}{Los Alamos National Laboratory, Los Alamos, NM 87545, USA}
\altaffiltext{4}{Universit\`a di Pisa and Istituto Nazionale di Fisica Nucleare, Sezione di Pisa I-56127 Pisa, Italy}
\altaffiltext{5}{Laboratoire AIM, CEA-IRFU/CNRS/Universit\'e Paris Diderot, Service d'Astrophysique, CEA Saclay, F-91191 Gif sur Yvette, France}
\altaffiltext{6}{Istituto Nazionale di Fisica Nucleare, Sezione di Trieste, I-34127 Trieste, Italy}
\altaffiltext{7}{Dipartimento di Fisica, Universit\`a di Trieste, I-34127 Trieste, Italy}
\altaffiltext{8}{Istituto Nazionale di Fisica Nucleare, Sezione di Padova, I-35131 Padova, Italy}
\altaffiltext{9}{Dipartimento di Fisica e Astronomia ``G. Galilei'', Universit\`a di Padova, I-35131 Padova, Italy}
\altaffiltext{10}{Istituto Nazionale di Fisica Nucleare, Sezione di Pisa, I-56127 Pisa, Italy}
\altaffiltext{11}{Istituto Nazionale di Fisica Nucleare, Sezione di Bari, I-70126 Bari, Italy}
\altaffiltext{12}{W. W. Hansen Experimental Physics Laboratory, Kavli Institute for Particle Astrophysics and Cosmology, Department of Physics and SLAC National Accelerator Laboratory, Stanford University, Stanford, CA 94305, USA}
% ...

\begin{abstract}
The \textit{Fermi} Large Area Telescope (LAT) has opened the way for comparative studies of cosmic rays (CRs) and high-energy objects in the Milky Way (MW) and in other, external, star-forming galaxies. Using 2 yr of observations with the \textit{Fermi} LAT, Local Group galaxy M31 was detected as a marginally extended gamma-ray source, while only an upper limit has been derived for the other nearby galaxy M33. We revisited the gamma-ray emission in the direction of M31 and M33 using more than 7 yr of LAT Pass 8 data in the energy range 0.1$-$100 GeV, presenting detailed morphological and spectral analyses. M33 remains undetected, and we computed an upper limit of $2.0\times 10^{-12}\rm \,erg\, cm^{-2}\,s^{-1}\,$ on the 0.1$-$100 GeV energy flux (95\% confidence level). This revised upper limit remains consistent with the observed correlation between gamma-ray luminosity and star-formation rate tracers and implies an average CR density in M33 that is at most half of that of the MW. M31 is detected with a significance of nearly $10\sigma$. Its spectrum is consistent with a power law with photon index $\Gamma = 2.4\pm0.1_{\rm stat+syst}$ and a 0.1$-$100 GeV energy flux of $(5.6\pm0.6_{\rm stat+syst})\times 10^{-12}\rm \,erg\, cm^{-2}\,s^{-1}$. M31 is detected to be extended with a $4\sigma$ significance. The spatial distribution of the emission is consistent with a uniform-brightness disk with a radius of $0\fdg4$ and no offset from the center of the galaxy, but nonuniform intensity distributions cannot be excluded. The flux from M31 appears confined to the inner regions of the galaxy and does not fill the disk of the galaxy or extend far from it. The gamma-ray signal is not correlated with regions rich in gas or star-formation activity, which suggests that the emission is not interstellar in origin, unless the energetic particles radiating in gamma rays do not originate in recent star formation. Alternative and nonexclusive interpretations are that the emission results from a population of millisecond pulsars dispersed in the bulge and disk of M31 by disrupted globular clusters or from the decay or annihilation of dark matter particles, similar to what has been proposed to account for the so-called Galactic center excess found in \textit{Fermi}-LAT observations of the MW.\\
\end{abstract}


\section{Introduction}

The gamma-ray luminosity of star-forming galaxies originates from the large-scale population of cosmic rays (CRs) interacting with the interstellar medium (ISM) and from the ensemble of discrete high-energy sources, such as supernova remnants, pulsars, and their nebulae, most of which result from the evolution of the short-lived and most massive stars. Additional contributions may come from an active galaxy nucleus or, more hypothetically, from the decay or annihilation of dark matter particles.

Studying the gamma rays of a galaxy whose emission arises predominantly from its star-formation activity can inform us about the acceleration of CRs in powerful objects and its transport through the ISM. Comparing different galaxies can then be a test of our understanding of these processes by revealing how global properties such as star formation rate (SFR), gas content and metallicity, or galaxy size affect the population of high-energy objects and CRs.

In the Milky Way (MW), diffuse interstellar emission dominates the gamma-ray output and has proven to be a rich source of information, even beyond the physics of CRs \citep[see the recent review by][]{Grenier15}. Besides the MW, \textit{seven} external star-forming galaxies have been firmly detected in gamma rays with the \textit{Fermi} Large Area Telescope (LAT), including the Large Magellanic Cloud \citep[LMC;][]{LMC10,LMC16}, the Small Magellanic Cloud \citep{SMC10}, the Andromeda galaxy M31 \citep[][hereafter Paper I]{M312010}, starburst galaxies M82 and NGC~253 \citep{M8210}, NGC~2146 \citep{NGC2146tang} and Arp 220 \citep{arp220Peng16,arp220Griffin16}. In Paper I, based on a subset of these detections, a correlation was suggested between gamma-ray luminosity and SFR; it was later strengthened by a large systematic study of more than 60 galaxies \citep{Ackermann12} and now appears as a possible constraint on the origin and transport of CRs \citep{Martin14}. More recently, a deep study of the LMC has shown that discrete sources can make up a significant contribution to the global gamma-ray output, especially the most exceptional ones \citep{LMCpsr15,corbet16}, and revealed extended emission with unexpected properties \citep{LMC16}; both findings confirm the need for more studies of external star-forming galaxies.

With their relatively high gas masses, star formation activities, and small distances to Earth, M31 and M33 have long been predicted to be gamma-ray sources. Earlier gamma-ray observations of M31 and M33 involved COS-B \citep{M31cosb} and EGRET \citep{M31EGRET1,M31EGRET2}, but only upper limits (ULs) were derived. Using 2 yr of LAT observations, Paper I reported a $5.3\sigma$ detection of M31 and a marginal spatial extension ($\sim 1.8\sigma$); at the same time, M33 was not detected, but it was suggested to be detectable within years if its gamma-ray luminosity obeys the above-mentioned correlation with SFR. As the only other large spiral in the Local Group of galaxies besides the MW, M31 is a highly relevant target for a comparative study. Moreover, with an angular size over 3$^\circ$, it is one of the rare nearby galaxies holding potential for a resolved analysis.

In this paper we revisited the gamma-ray emission from M31 and M33 using more than 7 yr of Pass 8 observations, which is the latest version of LAT data and has overall improved performance over previous Pass 7 data \citep{pass8Atwood}. The paper is organized as follows. We briefly introduce the \textit{Fermi}-LAT instrument and Pass 8 data in Section 2 and present in detail the morphological and spectral analysis in Section 3. We discuss possible interpretations of our findings in Section 4 and summarize our results in Section 5.

\section{Data set and analysis methods}

The LAT is a pair-conversion telescope comprising a $4 \times 4$ array of silicon strip trackers and cesium iodide calorimeters covered by a segmented anti-coincidence detector to reject charged-particle background events. The LAT covers the energy range from 20 MeV to more than 300 GeV with a field of view of 2.4 sr. It operates predominantly in survey mode and observes the entire sky every two orbits (3 hr) by rocking north and south about the orbital plane on alternate orbits \citep{LAT09}.

We used SOURCE class events, converting in both the front and back sections of the LAT, but excluding those with a zenith angle larger than $90^\circ$ or collected when the LAT's rocking angle was larger than $52^\circ$ to avoid the Earth limb contamination. We considered events with reconstructed energies in the energy range 0.1$-$100 GeV and with reconstructed directions within a $14^\circ \times 14^\circ$ region of interest (ROI). For the analysis of M31, we selected 88 months of Pass 8 data collected between 2008 August 4 and 2015 December 1, with an ROI center at  $(\alpha,\delta)=(10\fdg6847,41\fdg2687)$.  The data set used for the analysis of M33 spans 85 months, with an ROI centered on $(\alpha,\delta)=(23\fdg4621,30\fdg6599)$. The coordinates for both galaxies were taken from the SIMBAD\footnote{ http://simbad.u-strasbg.fr/simbad/} database and correspond to the J2000 epoch.

For each ROI, a complete spatial and spectral source model was built. We used the latest model gll\_iem\_v06.fits for the Galactic interstellar emission and the isotropic emission spectrum iso\_P8R2\_SOURCE\_V6\_v06.txt for the extragalactic emission and residual instrumental background. Point sources within $20^\circ$ around M31 or M33 in the LAT Third Source Catalog \citep[3FGL;][]{3FGL} were included in the model (except 3FGL~J0042.5+4117, which is M31), with spectral parameters set free to vary for sources within $5^\circ$ around M31 or M33. This source model not including M31 or M33 will hereafter be referred to as the background model. On top of this background model, we explored several possibilities for the  morphology and spectrum of M31 or M33.

Each source model was fitted to the data following a maximum likelihood approach for binned data and Poisson statistics \citep{mattox96}. Unless otherwise stated, we used a $0\fdg1 \times 0\fdg1$ pixel size and four logarithmic energy bins per decade. The analysis was performed using the P8R2\_SOURCE\_V6 Instrument Response Functions and the \fermi\ Science Tools version 10-01-01 available from the {\it Fermi} Science Support Center\footnote{ http://fermi.gsfc.nasa.gov/ssc/}. The significance of model components for M31 or M33 is quantified with the test statistic (TS), which is expressed as TS $=2(\log \mathcal{L}-\log \mathcal{L}_{0})$, where $\log \mathcal{L}$ and $\log \mathcal{L}_{0}$ are the logarithms of the maximum likelihood of the complete source model and of the background model (i.e. the source model without M31 or M33 included), respectively. The significance of the spatial extension of M31 or M33 is quantified by TS$_{\rm ext}$, which is twice the difference between the $\log \mathcal{L}$ obtained with an extended source model and that obtained with a point-like source model at its best-fit position.

As a potentially extended gamma-ray source, and one possibly shining because of interstellar processes, M31 requires some caution in the use of the Galactic interstellar emission model. This model is developed from radio and infrared tracers of interstellar gas \citep{catSNR16}. In particular, it is based on the Leiden$-$Argentine$-$Bonn 1.4\,GHz observations of atomic gas \citep{LAB05} and on a dust reddening map \citep{Schlegel98}, both of which are all-sky data in which M31 appears. M31 was removed from these maps in developing the interstellar emission model for the MW. Otherwise, any emission from M31 would be erroneously absorbed in the fitting of the Galactic interstellar emission model. In the case of the 1.4\,GHz data, M31 was removed by applying the following two cuts in the $(l,b,v_{\rm LSR})$ data space: (1) $l$ from $119^\circ$ to $123^\circ$, $b$ from $-23\fdg5$ to $-19\fdg5$, $v_{\rm LSR}$ up to $-120$\,km\,s$^{-1}$; and (2) $l$ from $121^\circ$ to $124^\circ$, $b$ from $-22^\circ$ to $-19\fdg5$, $v_{\rm LSR}$ from $-120$ to $-50$\,km\,s$^{-1}$. Examination of higher-resolution observations of this region from the Effelsberg$-$Bonn H\,{\sc i} survey \citep{Winkel16} confirms that such cuts effectively remove the great majority of the disk of M31 from the data. For $-30$\,km\,s$^{-1}$ $ > v_{\rm LSR} > $ $-50$\,km\,s$^{-1}$ (i.e. data not cut out), foreground emission from the MW blends with remaining signal from M31 at the northeastern tip of M31. We estimated that, on some lines of sight in this direction, up to $\sim40$\% of the signal from M31 might have been incorporated in the maps used in the Galactic interstellar emission model. Yet, this confusion happens over a very restricted region compared to the full extent of M31, and at a distance of $1\fdg25$ from the center of the galaxy, such that any gamma-ray emission correlated with the disk of M31 should safely be recovered. Another possible source of bias in the study of extended sources is that they may be part of the large-scale residuals reinjected into the final model \citep[see][for details]{catSNR16}, and we checked that it is not the case for the region around M31.

\section{Data analysis results}

\subsection{M31}

\subsubsection{Morphological analysis}
\begin{figure*}[p]
\centering
\includegraphics[scale=0.27]{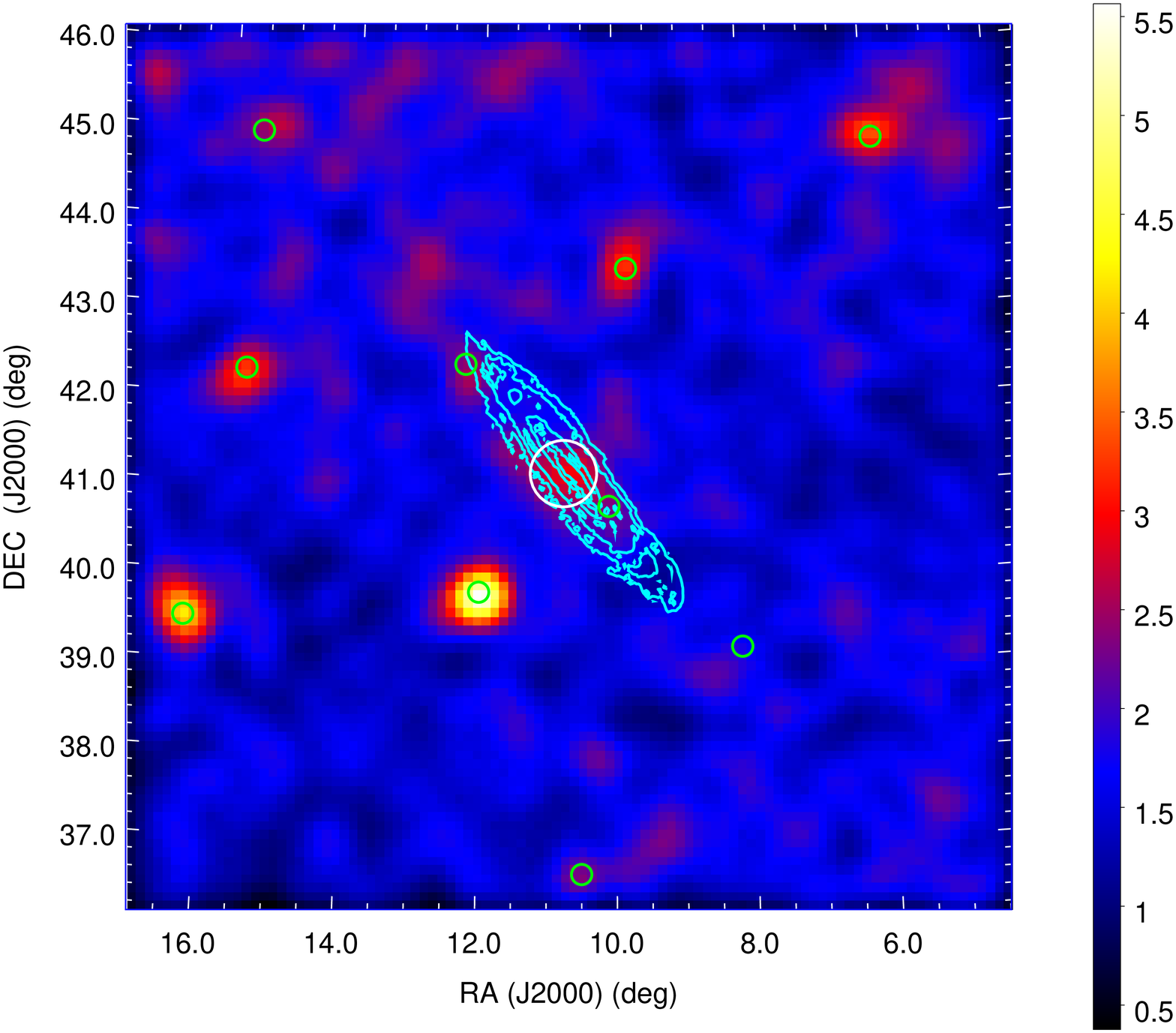}
\includegraphics[scale=0.27]{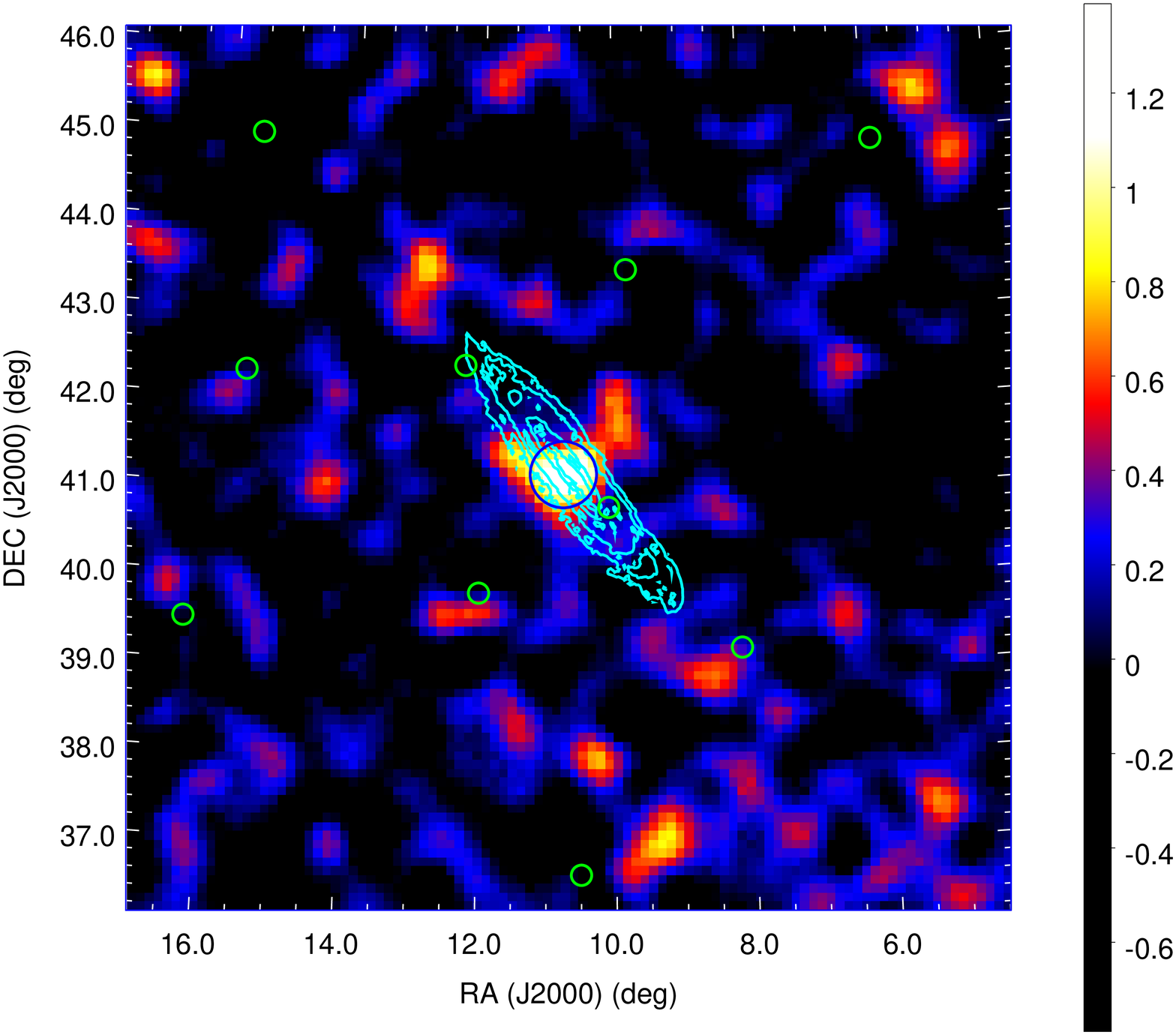} 
\caption{Counts map (left) and residual counts map after background subtraction (right), in units of counts pixel$^{-1}$, using 1$-$100 GeV events in a $10^\circ\times10^\circ$ region around M31. Overlaid are the best-fit disk model (white and blue circles for left and right panels, respectively), eight 3FGL point sources (green circles), and contours of the atomic gas column density map (cyan curves). Both maps have a pixel size of $0\fdg1$ and were smoothed with a Gaussian kernel with $\sigma=0\fdg4$.}
\label{cmap_m31}
\end{figure*}

Figure \ref{cmap_m31} shows the LAT counts map (left) and residual counts map after background subtraction (right) in the 1$-$100 GeV energy range. M31 is clearly visible in the counts map and appears more prominently in the residual map. The gamma-ray morphology of M31 is characterized using the $\mathtt{pointlike}$ tool \citep{Kerr10} on a data set restricted to energies above 1 GeV to benefit from the better angular resolution. We explored different geometrical models, such as point source, disk, elliptical disk, Gaussian, or elliptical Gaussian. We also considered spatial templates from observations at other wavelengths: \textit{Herschel}/PACS map at 160 $\mu m$, \textit{Spitzer}/IRAC map at 3.6 $\mu m$, and an atomic gas column density $N_{\rm H}$ map from \cite{Braun09}, uncorrected for self-opacity. The latter models are intended to test the spatial correlation of the gamma-ray emission with star formation sites, the old stellar population, or interstellar gas, respectively. We also tested two-component models such as a point source at the center of M31 and an extended component around it.
The spectrum of M31 was initially modeled by a simple power law (PL), an assumption that we revisited once a satisfactory spatial model is identified.
Fit results are reported in Table \ref{spatptlikeM31}.

Starting from a simple point-source model located at the center of M31, we found that optimizing the position of the point source provides a limited improvement with significance $< 2\sigma$, but allowing for an extension improves the fit with a significance below $3\sigma$ for the uniform-brightness disk model. Allowing for an offset of the disk center with respect to the center of M31 results in a slightly more significant extension and an offset from the center of M31 that is not significant ($<2\sigma$). Using a 2D Gaussian intensity distribution instead of a uniform-brightness disk degrades the fit likelihood by a negligible amount. Similarly, allowing for some elongations of the signal in some directions with elliptical disk or elliptical Gaussian models does not significantly improve the fit. Two-component models consisting of a point source and a disk or 2D Gaussian component around it, all centered at the M31 center, also led to very marginal improvements. These two-component models are therefore not required, especially since in each case the point-source component is not significantly detected.
 
Among template map models, the $N_{\rm H}$ map yields the fit with the lowest likelihood of all tested models. For the same number of degrees of freedom, the \textit{Herschel}/PACS or \textit{Spitzer}/IRAC maps are not favored compared to a simple point source at the center of M31, but the \textit{Spitzer}/IRAC map provides a slightly better fit to the data than the \textit{Herschel}/PACS map. These results are consistent with those obtained with geometrical models because the $N_{\rm H}$ map, and to a lesser extent the \textit{Herschel}/PACS map, are dominated by the relatively extended disk of M31, while the \textit{Spitzer}/IRAC map is dominated by its bulge. 

We retained the uniform-brightness disk with radius of $0\fdg38\pm 0\fdg05$ as the best-fit morphological model for M31 because it is the simplest of the best-fitting models. The different tests summarized above indicate that the emission is consistent with being symmetric around the center of the galaxy. Yet, we emphasize that, based on the data currently at our disposal, we cannot reject nonuniform-brightness distributions or multicomponent models. With such a disk model, M31 has TS $=51$ and TS$_{\rm ext}=7.6$ from an analysis in the 1$-$100 GeV band. Including lower-energy events down to 100 MeV results in a more significant detection and extension with TS $=95$ and TS $_{\rm ext}=16$ (and a source extension consistent with that obtained from the 1$-$100 GeV data analysis). In including lower-energy events, we verified that source 3FGL~J0040.3+4049 does not influence the results because of its proximity to M31 (the source lies within the optical or infrared disk of M31; see Figure \ref{cmap_m31}). In analyzing 1$-$100 GeV events, 3FGL~J0040.3+4049 is well resolved from M31 because it has a hard spectrum with photon index 1.3, but the poor angular resolution of the LAT below 1 GeV may introduce some cross-talk between both sources. Fixing the spectral parameters of 3FGL~J0040.3+4049 (either the spectral index only or both the index and the prefactor) to the values determined from the 1 to 100 GeV analysis yields TS $=97-98$ for M31, similar to the value obtained when parameters for 3FGL~J0040.3+4049 are left free in the fit, confirming that the source has little impact on the properties derived for M31.

\begin{figure*}[!htp]
\centering
\includegraphics[scale=0.28]{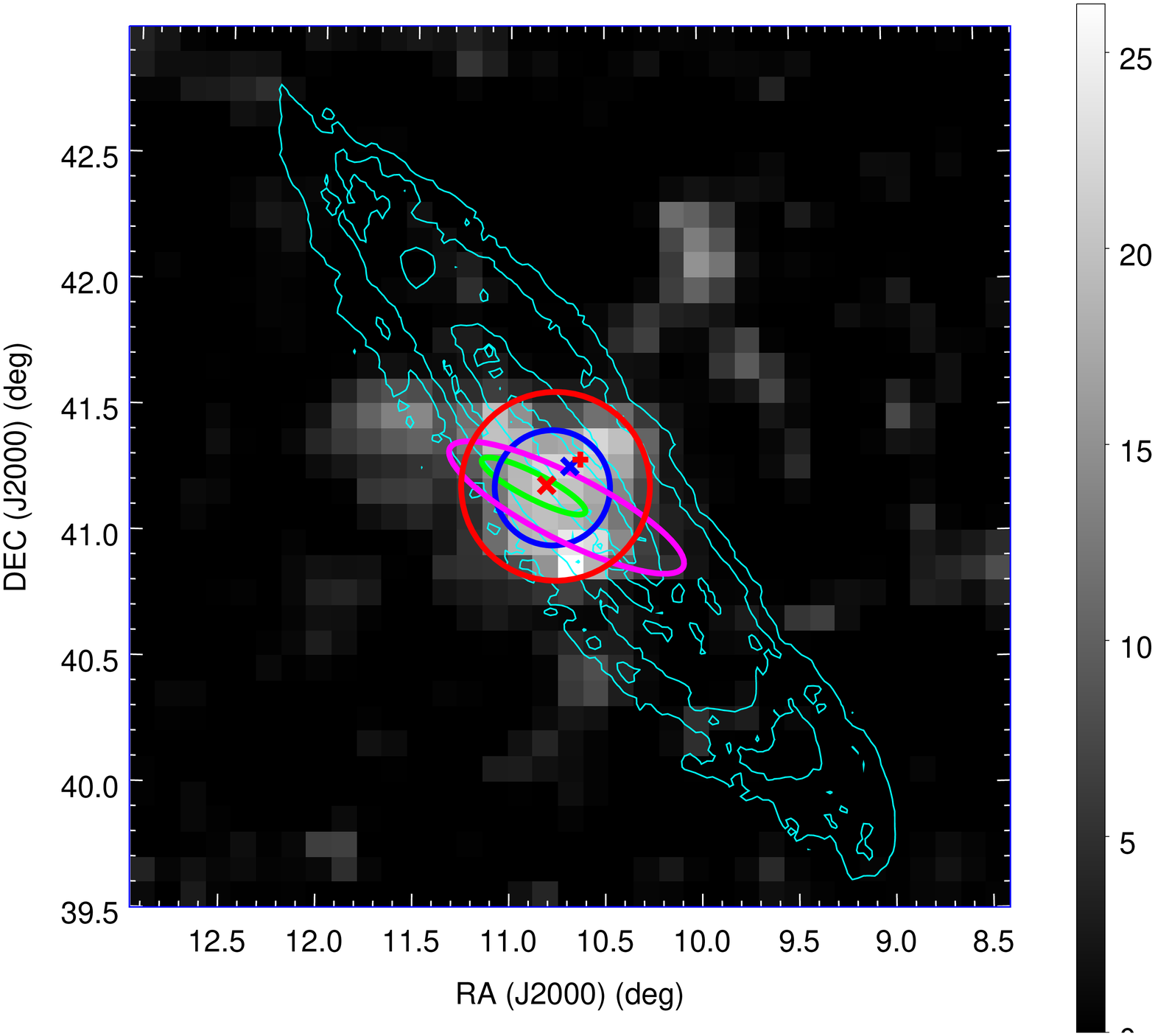} 
\includegraphics[scale=0.28]{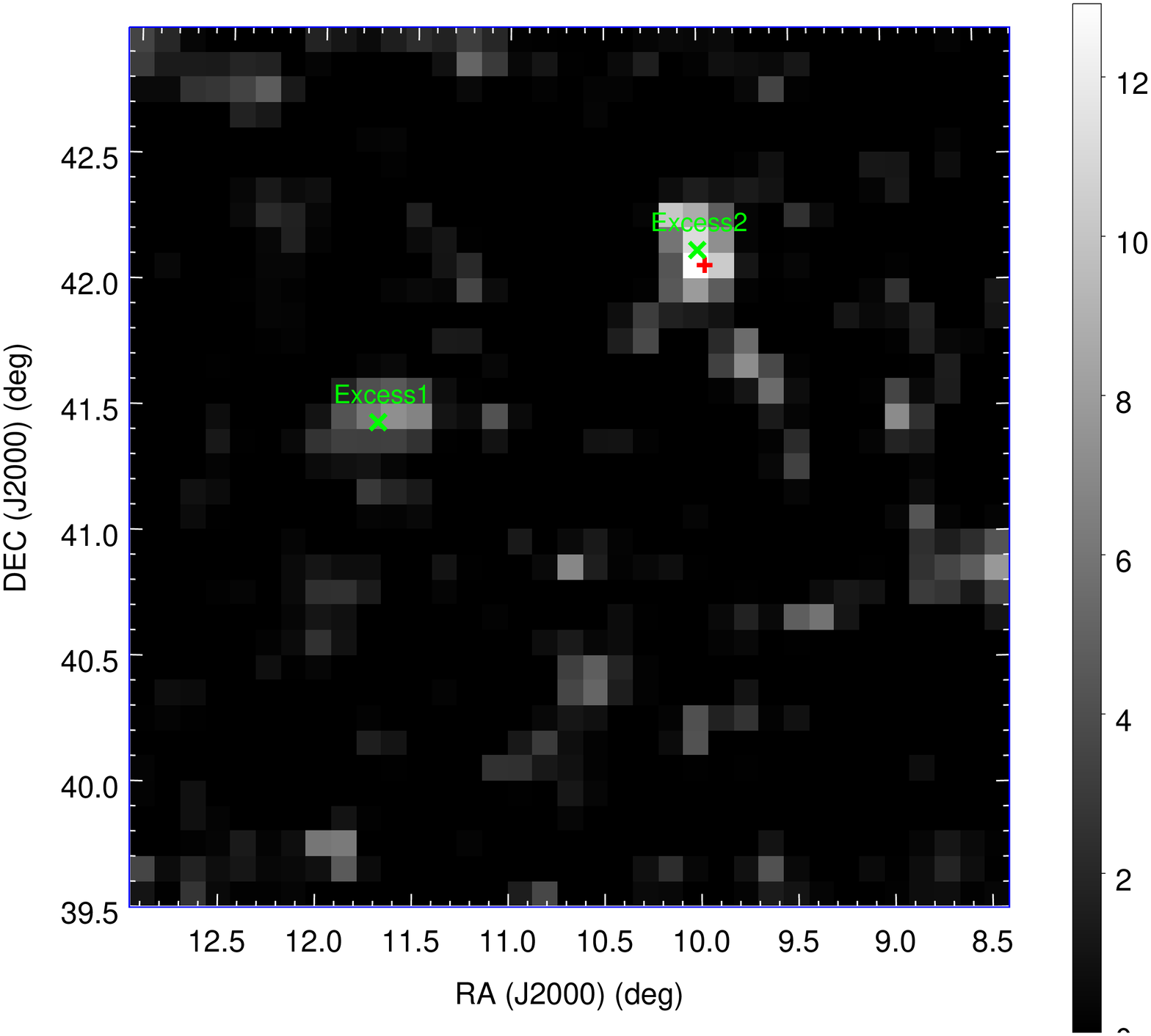} 
\caption{Left: TS map obtained for the background model, using 1$-$100 GeV events in a $3\fdg5\times3\fdg5$ region around M31. Overlaid are the 3FGL position of M31 (the red plus sign), the center of M31 from SIMBAD (the blue cross), the best-fit point source position (the red cross), the best-fit disk (the red circle), the best-fit elliptical disk (the magenta ellipse), the best-fit Gaussian (the blue circle, 1$\sigma$ extent), the best-fit elliptical Gaussian (the green ellipse, 1$\sigma$ extent), and contours of the atomic gas column density map (cyan curves). 
Right: TS map obtained for a source model including M31, using 1$-$100 GeV events in a $3\fdg5\times3\fdg5$ region around M31. Overlaid are the positions of possible sources Excess1 and Excess2 (the green cross) and of the LAT 7 yr internal list source (the red plus sign). Both maps have a pixel size of $0\fdg1$.}
\label{tsmap_m31}
\end{figure*}

Figure \ref{tsmap_m31} (left panel) shows the TS map for the background model, in a $3\fdg5 \times 3\fdg5$ region around M31 (that is, adding a point-source model to the background model and testing it over a grid of positions). Contours and shapes for the best-fit spatial models tested here are overlaid. This plot illustrates that the gamma-ray emission is clearly extended but over an area much smaller than the full extent of M31. The flux appears confined to the central parts of the galaxy and does not fill the disk or extend far from it. To investigate whether there are unmodeled emission components around M31 or whether multiple sources are necessary to account for the total emission in the direction of M31, we computed the TS map for a source model including M31. Figure \ref{tsmap_m31} (right panel) shows two residual point-like excesses to the east and northwest of M31. These were dubbed Excess1 and Excess2\footnote{Excess2 is spatially coincident with a source in the \textit{Fermi}-LAT Collaboration internal 7 yr source list, with an angular separation of $4.^{\prime}0$. }, and their optimal positions are given in Table \ref{spatptlikeM31}. They were added to our source model to evaluate their impact on the fit. With TS values of 8 and 12 for Excess1 and Excess2, respectively, both were below the standard detection threshold of 25. Comparing the $\log \mathcal{L}$ values of fits with and without components modeling the two excesses indicates a fit improvement with a significance $<3\sigma$, so we decided not to consider these components further in the analysis. We note, however, that including these sources in the background model results in the extension of M31 being smaller and less significant, suggesting that the emission might actually be even more confined to the inner regions than discussed above.

\newpage

\subsubsection{Spectral analysis}
\begin{figure*}[p]
\centering
\includegraphics[scale=0.45]{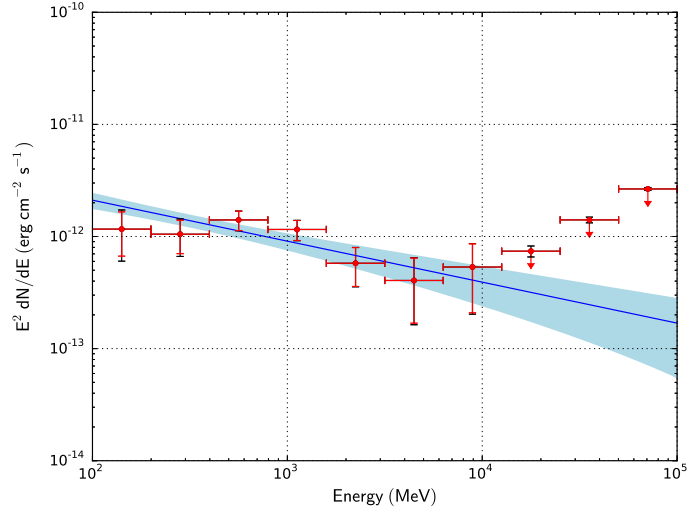} 
\caption{Spectrum of M31. The blue solid line is the best-fit PL model from an analysis over the full energy range, and the light-blue shaded area indicates the 68\% confidence level uncertainty domain. Red spectral points were obtained by performing independent fits in individual energy bins. Red arrows represent the 95\% confidence level flux ULs. Red and black vertical error bars are statistical and total uncertainties, respectively, with the latter being the quadratic sum of statistical and systematic uncertainties on the effective area.} 
\label{sed_m31}
\end{figure*}

For the spectral analysis of M31, we performed a binned maximum likelihood fitting in the 0.1$-$100 GeV energy range, using the $\mathtt{gtlike}$ tool provided in the \textit{Fermi} Science Tools, with 30 logarithmic energy bins in total. To avoid possible cross-talk, the spectral index of the background source mentioned above (3FGL~J0040.3+4049) was fixed to the value determined in the 1$-$100 GeV analysis, but we checked that leaving it free in this broadband analysis has a negligible impact. Using the best-fit disk model described above, we first compared a simple PL, a PL with exponential cutoff (PLEC), and a log-parabola (LP) for M31. The addition of a curvature in the spectrum does not significantly improve the fit ($<3\sigma$). The flux from M31 is satisfactorily described by a PL with photon index $\Gamma=2.4\pm0.1_{\rm stat+syst}$ and a 0.1$-$100 GeV energy flux of $(5.6\pm0.6_{\rm stat+syst})\times 10^{-12}\rm \,erg\, cm^{-2}\,s^{-1}$ (see below for the computation of systematic uncertainties). This is reported in Table \ref{specgtlike} and illustrated in Figure \ref{sed_m31}, where the best-fit PL model is plotted together with spectral points. The latter were determined by performing a maximum likelihood analysis in 10 logarithmically spaced energy bins over 0.1$-$100 GeV. Within each bin, the spectrum of M31 was modeled as a simple PL with fixed index $\Gamma=2$, and the normalization of M31 was allowed to vary while all other sources were fixed to their best-fit parameters obtained from the broadband analysis. ULs on the flux at 95\% confidence level were derived using the Bayesian method when M31 has TS $<4$ ($2\sigma$) in a given bin. For the spectral points and spectral parameters, systematic uncertainties in the LAT effective area were estimated by refitting the data using as a scaling functions to ``bracket'' the effective area, following the recommendations of the {\it Fermi} Science Support Center and using as a scaling function $\pm 5\%$ over 0.1$-$100 GeV. 
We checked that using a different photon index within each bin, e.g., 2.4 instead of 2.0, or setting normalizations free for diffuse components and sources within $2^\circ$ of M31, has an insignificant impact. In the latter test, the spectral parameters of 3FGL~J0040.3+4049 were fixed to those determined in the 1$-$100 GeV analysis, which tends to underestimate the uncertainties on the low-energy flux from M31 because the poor angular resolution of the LAT at low energies would have allowed some cross-talk between both sources. This choice is, however,  justified by the point-like nature and hard spectrum of 3FGL~J0040.3+4049.

To help pinpoint the possible origin of the emission from M31, we also tested more physically motivated spectral models (Table \ref{specgtlike}). We considered interstellar gamma-ray emission spectra from a GALPROP model of the MW \citep{Strong10}, selecting the plain diffusion model for a halo height of 4\,kpc, and an average spectrum of observed millisecond pulsars (MSPs) in the MW \citep{Cholis14}, which is a PLEC model with a photon index of 1.6 and a cutoff energy of 4 GeV. Still using the best-fit disk model, we found that the emission from M31 has a spectrum that is consistent with that of the total interstellar emission from the MW or with its pion-decay component (the $\log \mathcal{L}$ difference compared to the best-fit PL model is negligible); it is comparatively less consistent with an average MSP spectrum and with the inverse-Compton (IC) component of the interstellar emission from the MW (because the latter is too flat), but the differences in terms of $\log \mathcal{L}$ are modest. 

\subsubsection{Flux variability}
\begin{figure*}[!htp]
\centering
\includegraphics[scale=0.42]{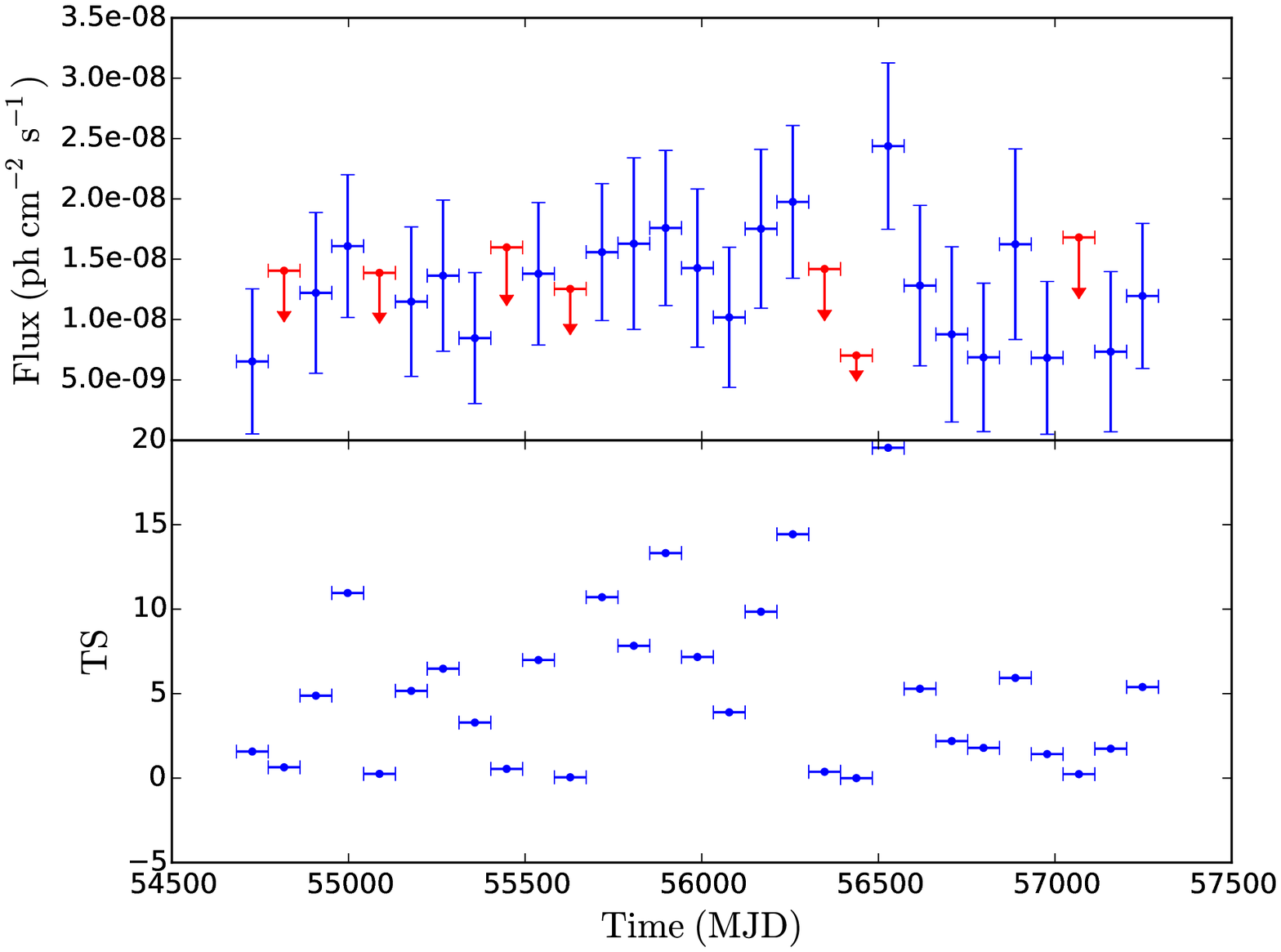} 
\includegraphics[scale=0.42]{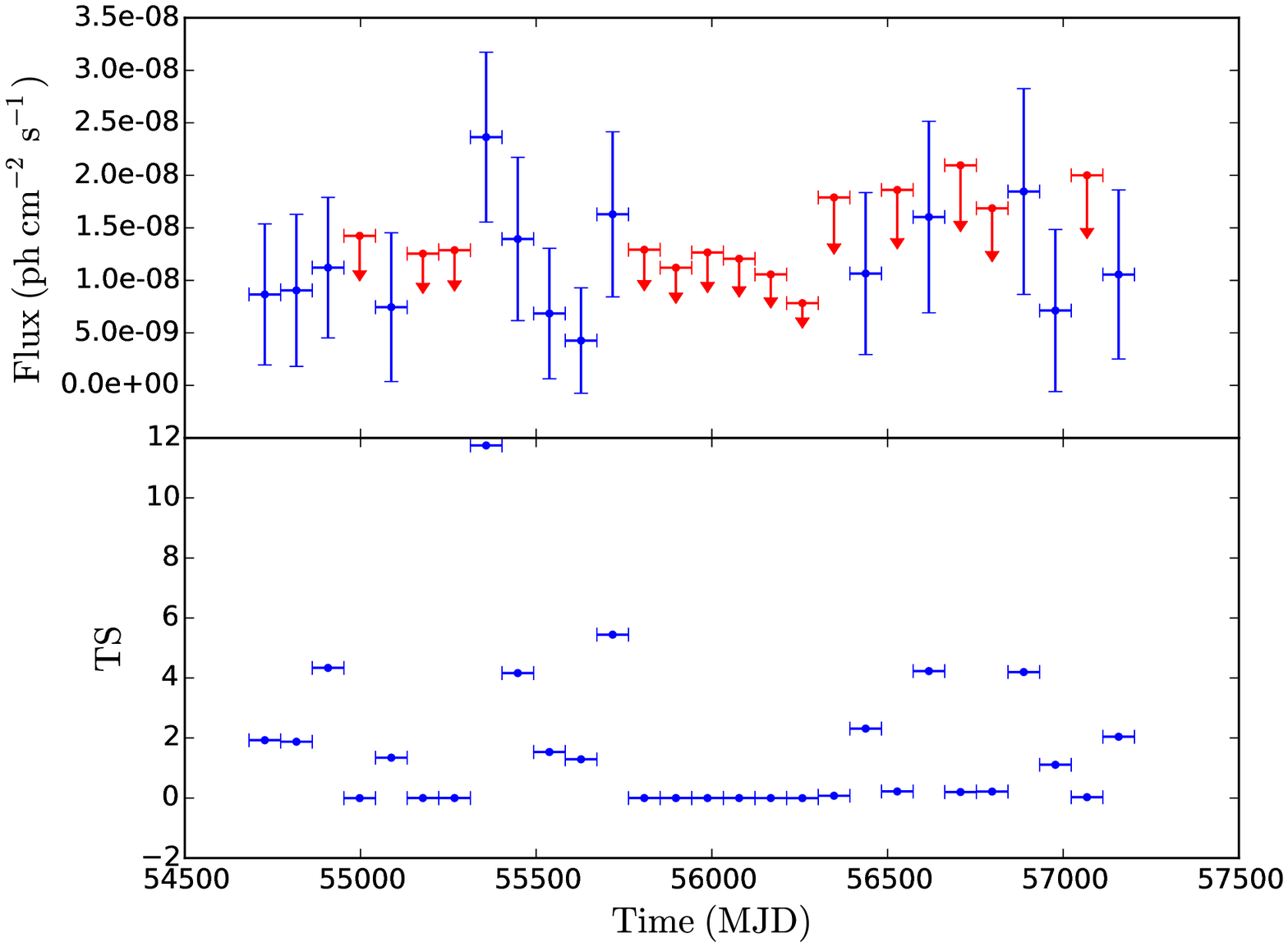} 
\caption{0.1$-$100 GeV light curve and TS evolution for M31 (left) and the source in the direction of M33 (right), with 90-day binning. Flux ULs at the 95\% confidence level are shown as red arrows in bins where the source has TS $<1$.}
\label{lc_M31M33}
\end{figure*}

To examine the variability of the gamma-ray flux from M31, we computed a long-term light curve with a 90-day binning, for events in the energy range 0.1$-$100 GeV. In each time bin, all sources (including M31) within $5^\circ$ of the nominal position of M31 had spectra fixed to the shapes obtained from the full data set analysis, and only normalizations were allowed to vary. ULs at 95\% confidence level were calculated when M31 had TS $<1$ in a given time bin. The result is shown in Figure \ref{lc_M31M33} (left panel). Using the 90-day binning, we quantified the variability significance following the same method used in \cite{3FGL} and obtained $1.4\sigma$ (for 28 degrees of freedom). The emission is therefore consistent with being steady, at least down to the scale of a few months.

\subsection{M33}

\subsubsection{Morphological analysis}
\begin{figure*}[!htp]
\centering
\includegraphics[scale=0.35]{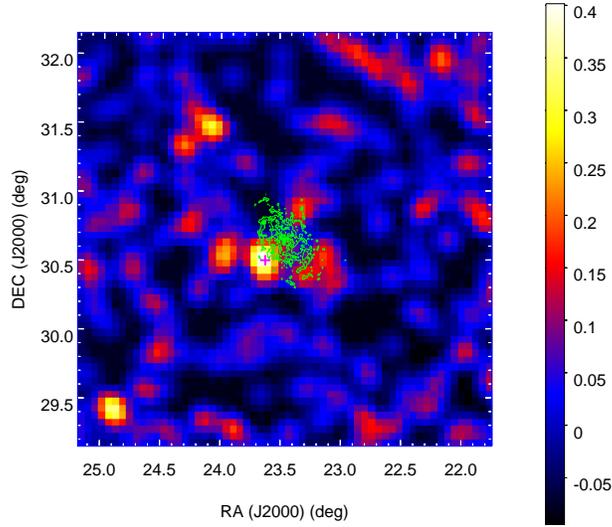} 
\caption{Residual counts map after background model subtraction, using 1$-$100 GeV PSF3 events in a $3^\circ\times3^\circ$ region around M33. Overlaid are the best-fit point source (the magenta plus sign) and contours of the \textit{Herschel}/PACS map at 160 $\mu m$ (green curves). The map has a pixel size of $0\fdg05$ and was smoothed with a Gaussian kernel with $\sigma=0\fdg3$.}
\label{cmap_m33}
\end{figure*}

We repeated the procedure used for M31 to characterize the gamma-ray morphology of M33. Using SOURCE class data and event energies $>1$ GeV, weak excess emission appears in the direction of M33, but at a level that seems comparable to other positive fluctuations in the field. The excess is consistent with a point-like source with TS $=8$, at a position that is slightly offset from M33. To establish whether this weak source may be spatially associated with M33, we restricted the data set to the PSF3 subclass events, which have the most accurately reconstructed directions. Figure \ref{cmap_m33} shows the corresponding residual counts map after background subtraction.

We explored different spatial models for M33, and the result is that the gamma-ray emission in the direction of M33 is consistent with a point-like source at position $(\alpha,\delta) =(23\fdg625\pm0\fdg047, 30\fdg509\pm0\fdg043)$, $0\fdg2$ offset from the center of M33 (Table \ref{spatptlikeM33}). The TS of the source is 23 for an analysis in the 1$-$100 GeV range, and 28 when including lower-energy events down to 100 MeV. Spatial models consisting of a \textit{Herschel}/PACS map at 160 $\mu m$ or of a point source at the center of M33 can be excluded at the $\geq 3\sigma$ confidence level. 

\begin{figure*}[!htp]
\centering
\includegraphics[scale=0.35]{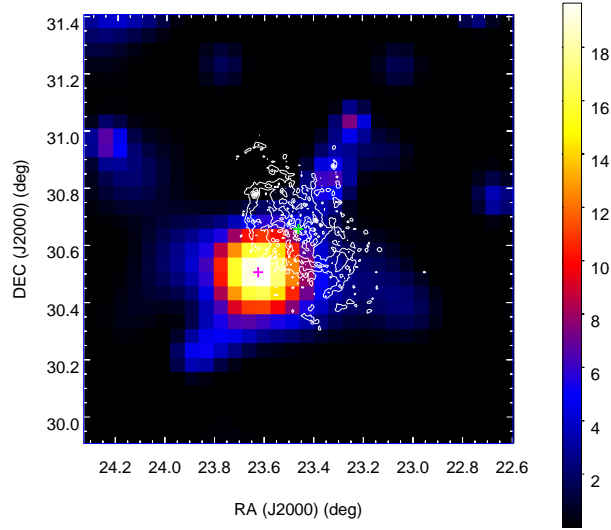} 
\caption{TS map obtained for the background model, using 1$-$100 GeV PSF3 events in a $1\fdg5\times1\fdg5$ region around M33. Overlaid are the M33 infrared center (the green plus sign), the best-fit point source (the magenta plus sign), and contours of the \textit{Herschel}/PACS map at 160 $\mu m$ (white curves). The map has a pixel size of $0\fdg05$.} 
\label{tsmap_m33}
\end{figure*}

Figure \ref{tsmap_m33} shows a TS map for the background model and events energies $>1$ GeV.
Overlaid are the position of the center of M33, the best-fit point-source position and the \textit{Herschel}/PACS map contours. 
The plot illustrates that the gamma-ray emission is most likely not connected to M33. The source may be a background active galaxy nucleus. To evaluate this possibility, we searched for variability of the signal but found nothing significant (see Section 3.2.3).

\subsubsection{Spectral analysis}
We performed a spectral analysis using $\mathtt{gtlike}$, for all SOURCE events in the energy range 0.1$-$100 GeV. 
Since we concluded above that the weak gamma-ray emission in the direction of M33 is not connected to the latter, we computed a flux UL for M33 over the entire energy band. In addition to the point source offset from M33, M33 was included in the source model either as a point source or as the \textit{Herschel}/PACS map template, and we assumed a PL spectrum with a fixed index of 2.2 (the typical value found for other detected star-forming galaxies; see \citealt{Ackermann12}). Using the \textit{Herschel}/PACS template, we obtained a photon flux UL of $0.3\times 10^{-8}\rm \,ph\, cm^{-2}\,s^{-1}\,$ and an integrated energy flux UL of $2.0\times 10^{-12}\rm \,erg\, cm^{-2}\,s^{-1}\,$ (95\% confidence level).

\subsubsection{Flux variability}

To examine the variability of the gamma-ray flux from the source in the direction of (but offset from) M33, we computed a long-term light curve with a 90-day binning, for all SOURCE events in the energy range 0.1$-$100 GeV. We followed a procedure similar to that used for M31 in Section 3.1.3 and the result is shown in Figure \ref{lc_M31M33} (right panel). For 27 degrees of freedom, the gamma-ray emission from the source is consistent with a constant signal, with a variability significance of 1.2$\sigma$.

\section{Discussion}

\subsection{Update to the $L_{\gamma}-$SFR correlation}

We derived for M31 and M33 the gamma-ray luminosity $L_{\gamma}=4\pi d^2 F_{100}$, where $d$ is the distance to the galaxy and $F_{100}$ is the photon flux above 100\,MeV. Assuming that the 0.1$-$100 GeV gamma-ray emission is dominated by CRs interacting with interstellar gas (via pion production and decay and to a smaller extent bremsstrahlung), one can compute an average emissivity per hydrogen atom as $\overline{q}_\gamma=L_{\gamma}/N$, where $N=1.19 \times 10^{57} \times (M_{\rm HI}+M_{\rm H_2})$ is the total number of hydrogen atoms in a galaxy, with $M_{\rm HI}$ and $M_{\rm H_2}$ being the mass of neutral and molecular hydrogen, respectively, in $\rm M_\odot$ units, and the conversion factor is in H-atom$\rm/ M_\odot$. The input parameters and results are summarized in Table \ref{emissM31M33}.
 
For M31, our $F_{100}$, $L_{\gamma}$, and $\overline{q}$ estimates are completely consistent with those reported in Paper I. For M33, we get ULs on $L_{\gamma}$ and $\overline{q}$ that are $\sim40\%$ lower than those in Paper I. The mildly improved constraints on M33 do not challenge the observed $L_{\gamma}-$to$-$SFR correlation \citep{Ackermann12}, especially because M33 was lying on the upper side of the estimated intrinsic dispersion that spans a bit less than an order of magnitude in $L_{\gamma}$ \citep[such a large scatter is expected from modeling of the interstellar emission from star-forming galaxies; see][]{Martin14}. Assuming that their gamma-ray emission results from CR$-$gas interactions, the inferred emissivities suggest that M31 and M33 have an average CR density that is at most half that of the MW. 

\begin{figure*}[!htp]
\centering
\includegraphics[scale=0.5]{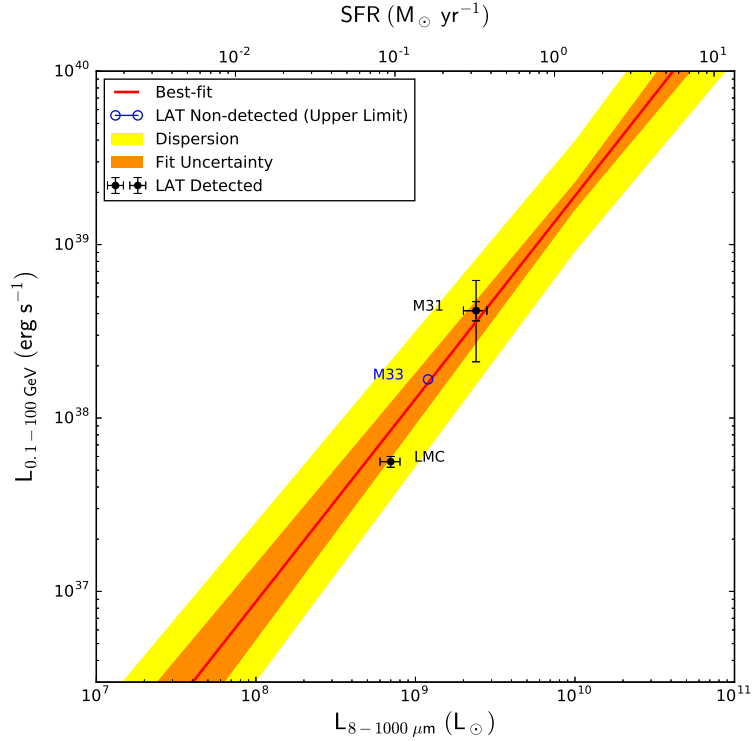} 
\caption{Gamma-ray luminosity (0.1$-$100 GeV) vs. total infrared luminosity (8$-1000\ \mu m$) plot from Figure 4 of \cite{Ackermann12}. The best-fit PL relation obtained from the study of 69 star-forming galaxies is shown by the red line, along with the fit uncertainty (darker shaded region) and intrinsic dispersion around the fitted relation (lighter shaded region). Updated measurements for M31 and the LMC are indicated as filled black circles, and the revised UL for M33 is shown as an open blue circle. The M31 point has a double error bar in ordinate; the smaller one corresponds to the quadratic sum of statistical and systematic uncertainties on the luminosity of the detected source, while the larger one represents the uncertainty on the gas-related contribution to the signal (see Section 4.1). The upper abscissa shows the estimated SFR from the infrared luminosity according to \cite{Kennicutt98}.}
\label{Lg_LIR}
\end{figure*}

Yet, in the case of M31, this assumption of emission being from CR$-$gas interactions can now be questioned. The observed emission is concentrated within a $0\fdg4$ angular radius, which translates into a physical radius of about 5\,kpc at the distance of M31. Most of the atomic and molecular gas in M31 actually lies beyond this radius, in a ring located at 10\,kpc and beyond it (see, e.g., \citealt{Smith12}). This extended gas ring is also where most of the star formation occurs \citep{Ford13}, and consequently where most sources of CRs such as supernova remnants are supposed to be. Yet, we did not detect such an extended contribution to the signal. Using the best-fit disk model, we derived a 95\% confidence level UL of $0.5\times 10^{-8}\rm \,ph\, cm^{-2}\,s^{-1}\,$ for additional 0.1$-$100 GeV emission correlated with the gas disk of M31 (as traced by the $N_{\rm H}$ map from \citealt{Braun09}). The average gas-related contribution to the emission from M31 thus has to be smaller than 50\% of the currently estimated flux of the galaxy. Depending on the nature of this observed central emission in M31 (interstellar or not; see below), the total interstellar luminosity could therefore be up to 50\% higher or more than 50\% lower than previously assumed. In either case, this does not challenge the observed $L_{\gamma}-$to$-$SFR correlation for the reasons given above. This is illustrated in Figure \ref{Lg_LIR}, where we used the correlation from Figure 4 of  \cite{Ackermann12}, along with our updated measurements for M31 and M33 and the revised estimate for the diffuse emission from the LMC presented in \cite{LMC16}.

\subsection{Interstellar emission}

If the observed central emission of M31 is interstellar in origin, it can be accounted for in at least two ways.

A first possibility is that the emission is gas related and the low gas content of the area subtended by the gamma-ray emission is compensated by a higher CR density, and hence gas emissivity, in the inner regions of M31. Yet, this relatively high density of CRs would be found several kiloparsecs away from the main sites of current or recent star formation \citep{Ford13}, while these sites of star formation do not shine in gamma rays at a detectable level despite being gas-rich. This is reminiscent of a discussion of the gamma-ray emission of the LMC \citep{LMC16}, in which areas relatively devoid of gas and star formation were found to be sites of significant gamma-ray production.

A second possibility is that the emission is dominated by IC scattering of a population of energetic electrons in the dense radiation field in the inner regions of M31 resulting from the large concentration of stars. In Section 3.1.2, we showed that the measured gamma-ray spectrum is slightly more consistent with a pion-decay spectrum than with an IC spectrum; on the other hand, we used models of the MW for these spectral fits, and M31 may have a different IC spectrum because of a different interstellar radiation field. The required central population of energetic electrons is not dominating in a radio synchrotron map of M31 (see Figure 2 of \citealt{Taba13}), where strong synchrotron emerges from the center of the galaxy but on a much smaller scale than that of the observed gamma-ray emission. Yet, synchrotron emission also depends on the distribution of the magnetic field in the galaxy, so the same population of energetic electrons may show up differently in synchrotron and IC. A puzzling fact is that IC emission would thus dominate the gas-related emission. In MW models, IC amounts to at most 45\% of the luminosity of the gas-related components, and such a high fraction implies a large confinement volume (see Table 2 of \citealt{Strong10}); it is not straightforward to figure out why a large galaxy like M31 would exhibit the opposite relation, i.e., gas-related emission being at most 50\% of the IC emission. 

A possible solution to these apparent discrepancies is that those high-energy particles responsible for the gamma-ray emission in the inner regions of M31 are not CRs resulting from recent star formation activity. The latter is thought to be the dominant source of non-thermal particles in the MW (energetically speaking), and this is assumed in the MW models referred to. In M31, however, because of its 10 times lower SFR compared to the MW \citep{Ford13}, the population of energetic particles in the inner regions may be contributed for the most part by another source, an old stellar population (see below) or the central supermassive black hole, for instance.

\subsection{Unresolved source population}

An alternative scenario is that the emission is not interstellar in origin but comes from a population of unresolved objects. The lack of correlation with the distribution of star formation sites does not favor sources related to short-lived massive stars, such as supernova remnants or normal pulsars and their nebulae. Instead, the location of the emission in the inner regions of M31, where a significant fraction of the old stellar population can be found \citep{Barmby06}, and where the largest concentration of X-ray sources is \citep{Voss07,Stiele10}, supports low-mass X-ray binaries and/or MSPs as possible sources of the signal. 

Such a situation is reminiscent of discussions about the nature of the so-called Galactic center (GC) excess \citep[see e.g.,][]{Abazajian12,Gordon13,Mirabal13,Yuan14}. In particular, \cite{Brandt15} suggested that a population of MSPs deposited in the MW inner regions by the disruption of globular clusters can account for all observed properties of the GC excess. This scenario implies a deposited stellar mass of about $5 \times 10^8 \, \rm M_\odot$, in a central region extending out to a galactocentric radius of 10 kpc \citep{Gnedin10}, associated with an average flux at 2 GeV of $2 \times 10^{-15} \rm \,GeV\, cm^{-2}\,s^{-1}\,$ per unit deposited stellar mass, at a distance of 8.3\,kpc \citep{Brandt15}. This translates into a total flux of $10^{-6} \rm \,GeV\, cm^{-2}\,s^{-1}\,$. For comparison, the flux at 2 GeV from M31 translated to a distance of 8.3\,kpc is $4 \times 10^{-6} \rm \,GeV\, cm^{-2}\,s^{-1}\,$ and the bulk of the emission comes from within a radius of 5\,kpc. The $\sim$4 times higher flux in M31 can be attributed to the number of globular clusters being 3--4 times greater in M31 than in the MW \citep{Galleti07}, which could result from a proportionately higher initial mass in globular clusters that subsequently dissolved in the disk and bulge of M31.


In Section 3.1.2, we showed that an average MSP spectrum is almost as good a fit to the data as a PL with free parameters, so the interpretation of the emission from M31 being due to populations of MSPs cannot be rejected from spectral arguments. Moreover, the lack of a significant curvature in the observed spectrum can be the result of a possible additional contribution to the signal from IC emission by the pairs released by the pulsars in the ISM \citep{Petrovic15}. The observed PL spectrum for M31 differs from that inferred for the GC excess; it is flat in the 0.1$-$1 GeV range, while the GC excess spectrum seems to cut off below 1 GeV. Yet, at these energies, the point-spread function of the LAT becomes relatively large (68\% containment radius above 1$^\circ$), and the derivation of the GC excess spectrum is affected by large uncertainties \citep[][submitted]{GCE_fermi}. In that respect, our external vantage point on M31 may provide a cleaner view of the central emission from a grand-design spiral galaxy, and in particular the contribution of old stellar populations: M31 has a 10 times lower SFR than the MW \citep{Ford13}, which should decrease the disk emission, while its bulge is $5-6$ times more massive \citep{Tamm12,Licquia15}, which could enhance any contribution from old objects.

\subsection{Dark matter}

Another possible interpretation of the central, extended, and seemingly symmetric emission from M31 is that it results from the decay or annihilation of dark matter particles. To evaluate whether such an interpretation is likely, we made a naive estimate of the expected signal from dark matter and compared it to the measured value. The calculation involves so-called \textit{J}-factors that were computed for Navarro$-$Frenk$-$White distributions of the smooth dark matter halo component \citep{Navarro97}.

Using the GC excess as a reference (but emphasizing that the interpretation of the latter in terms of dark matter is far from obvious), a flux at 2 GeV of $2\times10^{-7} \rm \,GeV\, cm^{-2}\,s^{-1}$ is measured, and the \textit{J}-factor over the studied region is $2\times 10^{22} \rm \,GeV^2\, cm^{-5}$ \citep[][submitted]{GCE_fermi}. In M31, the \textit{J}-factor integrated over the extent of the detected gamma-ray signal is $8\times 10^{18} \rm \,GeV^2\, cm^{-5}$ \citep[][in preparation]{DM_fermiHAWC}, but this value should be considered as uncertain by a factor of a few because the lack of rotation curve data within 7 kpc of the center of M31 results in large uncertainties in the central density distribution \citep{Tamm12}. From the ratio of \textit{J}-factors, one would expect a flux at 2 GeV from dark matter annihilation or decay in M31 of $8\times 10^{-11} \rm \,GeV\, cm^{-2}\,s^{-1}$, which is a factor 5 below the observed value. Because of the uncertainties in the \textit{J}-factor estimates, we cannot exclude from simple photometric arguments the possibility that dark matter accounts for a significant fraction of the observed signal. A dedicated analysis to characterize the dark matter contribution to the gamma-ray signal of M31 is beyond the scope of this paper and will be presented elsewhere \citep[][in preparation]{DM_fermiHAWC}.

\section{Conclusion}

We have analyzed more than 7 yr of \textit{Fermi}-LAT Pass 8 0.1$-$100 GeV observations of the Local Group galaxies M31 and M33. M33 is still undetected, and the flux UL we derived is $\sim40\%$ lower than that determined in 2010 from 2 yr of Pass 6 data. In contrast, M31 is detected with a significance of nearly $10\sigma$. The main improvement compared to our previous analysis is that gamma-ray emission from M31 is now detected as extended. This extension, however, is rather limited, and consequently its significance remains modest, at the $4\sigma$ level. The spatial distribution of the signal is consistent with a uniform-brightness disk with an angular radius of $0\fdg4$, 5\,kpc at the distance of M31, and no offset from the center of the galaxy, but nonuniform or multicomponent intensity distributions cannot be dismissed based on the current observations. The small extent of the source seems to exclude emission coming from the main gas ring and from the dominant star formation sites, contrary to expectations for typical interstellar emission. Possible and nonexclusive interpretations include a population of unresolved sources, energetic particles originating in sources not related to massive star formation, or dark matter. This result should be helpful in clarifying the origin of the excess gamma-ray emission observed in the inner regions of our Galaxy.

\vspace{0.3cm}

The \textit{Fermi} LAT Collaboration acknowledges generous ongoing support from a number of agencies and institutes that have supported both the development and the operation of the LAT, as well as scientific data analysis. These include the National Aeronautics and Space Administration and the Department of Energy in the United States; the Commissariat \`a l'Energie Atomiqueand and the Centre National de la Recherche Scientifique/Institut National de Physique Nucl\'eaire et de Physique des Particules in France; the Agenzia Spaziale Italiana
and the Istituto Nazionale di Fisica Nucleare in Italy; the Ministry of Education, Culture, Sports, Science and Technology (MEXT), High Energy Accelerator Research Organization (KEK), and Japan Aerospace Exploration Agency (JAXA) in Japan; and the K.~A.~Wallenberg Foundation, the Swedish Research Council, and the Swedish National Space Board in Sweden. Additional support for science analysis during the operations phase is gratefully acknowledged from the Istituto Nazionale di Astrofisica in Italy and the Centre National d'\'Etudes Spatiales in France.

X.H. is supported by the National Natural Science Foundation of China through grant 11503078 and by the Ministry of Science and Technology of the Republic of China (Taiwan) through grant 104-2811-M-007-059. A.K.H.K. is supported by the Ministry of Science and Technology of the Republic of China (Taiwan) through grant 103-2628-M-007-003-MY3. J.C.W. and M.Z. are supported by the National Natural Science Foundation of China through grant 11573060.

The authors thank Pauline Barmby for providing the IRAC 3.6 $\mu$m data and Annie Hughes and Robert Braun for providing the gas column density data. P.M. thanks Amaury Fau for his early works on the subject.

This research has made use of the SIMBAD database, operated at CDS, Strasbourg, France.

\bibliographystyle{apj}
\bibliography{m31}


\begin{sidewaystable}    
\begin{center}
\caption{\small Morphological 1$-$100 GeV fit results for M31.}
\begin{tabular}{lcccccccccc}
\toprule %
Spatial Model  & TS  &TS$_{\rm ext}$  &$-\log \mathcal{L}$ &R.A. (deg)  &Decl. (deg)  &Radius (deg) & Major Axis (deg) & Minor Axis (deg) & Position Angle (deg) & N$_{\rm dof}$ \\
  &  &  & &   &   &  &  &   & & \\
\hline %
Point source (fixed) & 41 &... & 193023.9   &10.6847 &41.2687  &... &... &... &... &2\\
Point source  (free) & 44 &... & \hphantom{193}022.4   &$10.81\pm0.07$ &$41.19\pm0.05$  &... &... &... &... &4\\
Disk (free center) & 51 &7.6 & \hphantom{193}018.6 &$10.76\pm0.06$ &$41.19\pm0.04$ &$0.38\pm0.05$ &... &... &... &5 \\
Disk (fixed center)  & 48 & 7.0 & \hphantom{193}020.4 &10.6847 &41.2687 &$0.39\pm0.06$ &... &... &... &3 \\
Elliptical disk &51 & 7.0 & \hphantom{193}018.9 &$10.70\pm0.09$ &$41.11\pm0.04$ &... &$1.05\pm0.16$ &$0.26\pm0.03$ &$63 \pm 2$ &7\\
Gaussian &50 & 6.2 & \hphantom{193}019.3 &$10.78\pm0.08$ &$41.18\pm0.07$ &$0.23\pm0.08$  &... &... &... &5 \\
Elliptical Gaussian  &51 &7.6 & \hphantom{193}018.6 &$10.88\pm0.09$ &$41.19\pm0.06$ &... &$0.11\pm0.08$ &$0.46\pm0.14$ &$-28\pm8$ &7\\
\textit{Herschel}/PACS map &36 &... & \hphantom{193}026.0 &... &... &... &... &... &... &2 \\
\textit{Spitzer}/IRAC map &42 &... & \hphantom{193}023.2 &... &... &... &... &... &... &2 \\
$N_{\rm H}$ map &26 &... & \hphantom{193}031.3 &... &... &... &... &... &... &2 \\
\hline %
\multicolumn{11}{c}{Multicomponent Model 1}\\
\hline %
Point source+disk &19, 11 &...    & \hphantom{193}018.4  &10.6847 &41.2687&$0.90\pm0.15$  &... &... &... &5 \\
Point source+Gaussian &14, 11 &... & \hphantom{193}018.7  &10.6847 &41.2687  &$0.50\pm0.18$ &... &... &... &5  \\
Point source+$N_{\rm H}$ map  &28, 4 &... & \hphantom{193}021.7  &10.6847 &41.2687  &... &... &... &... &4  \\
\hline
\multicolumn{11}{c}{Multicomponent Model 2}\\
\hline %
Disk &43 &6.1 &   &$10.74\pm0.06$  &$41.18\pm0.05$  &$0.36\pm0.05$ &...  &... &... & \\
Excess1 &8 &... & \hphantom{193}008.7   &$11.70\pm0.09$ &$41.44\pm0.07$  &... &... &... &... &13 \\
Excess2 &12 &... &   &$10.00\pm0.11$ &$42.13\pm0.05$  &...  &... &... &... & \\
\hline 
\label{spatptlikeM31}
\end{tabular}
\end{center}
Notes: Uncertainties are statistical only. The first three digits of the $-\log \mathcal{L}$ values are the same for all runs presented in this table, so we omitted them after the first line for readability. The epoch for the coordinates is J2000.
\end{sidewaystable}

\begin{table*}[h]
\begin{center}
\caption{\small Spectral  0.1$-$100 GeV fit results for M31 and M33.}
\begin{tabular}{lcccccc}
\toprule %
Model   &TS   &$-\log \mathcal{L}$  &$\Gamma$  & $E_{\rm cut}$ & $F_{100}$  & $G_{100}$   \\
(spatial$-$spectral) &   &  &  & (GeV) &($10^{-8}\rm ph\, cm^{-2}\,s^{-1}$)  & ($10^{-12}\rm erg\, cm^{-2}\,s^{-1}$)   \\
\hline %
\multicolumn{7}{c}{M31}\\
\hline %
Disk-PL   & 97 & 256909.9  &$2.4\pm0.1$   &...  &$1.0\pm0.2$  &$5.6\pm0.6$  \\
Disk-PLEC   & 99 & \hphantom{256}908.9  &$2.1\pm0.2$   & $5.3 \pm 4.9$ &$0.9\pm0.2$  &$4.8\pm0.7$  \\
Disk-LP   & 100 & \hphantom{256}908.6  &$2.4\pm0.1$ / $0.15\pm0.12$  &... &$0.8\pm0.2$  &$4.7\pm0.8$  \\
Disk-$\pi^0$  & 94  & \hphantom{256}911.5  & ...  &... & $0.5\pm0.1$ &  $4.3\pm0.5$\\
Disk-IC  & 92  & \hphantom{256}912.7 &...   &... &$0.7\pm0.1$  & $5.4\pm0.6$  \\
Disk-MW  & 96  & \hphantom{256}910.6   &...  &... &$0.6\pm0.1$ &$4.7\pm0.5$ \\
Disk-MSP   & 89 & \hphantom{256}914.6   & 1.6 (fixed)   & 4 (fixed) &$0.4\pm0.1$ &$3.8\pm0.5$\\
\hline
\multicolumn{7}{c}{M33}\\
\hline %
Point source-PL   &1 &  & 2.2 (fixed)   & &$<0.2$ &$<1.7$   \\
\textit{Hersche}l map-PL    &3    & &2.2 (fixed)   & &$<0.3$ &$<2.0$   \\
\hline 
\label{specgtlike}
\end{tabular}
\end{center}
Notes: PL stands for power law, PLEC for power law with exponential cutoff, and LP for log-parabola; MW designates the interstellar emission spectra from a GALPROP model of the MW, and $\pi^0$ and IC are its pion-decay and inverse-Compton components, respectively; MSP is the average observed spectrum of millisecond pulsars in the MW. 
In the case of the LP spectrum, the $\Gamma$ column contains the two indices (usually denoted $\alpha$ and $\beta$). $F_{100}$ and $G_{100}$ are the photon flux and energy flux above 100\,MeV, respectively. Only statistical uncertainties are shown, except for the PL model for M31, in which case they were quadratically added to systematic uncertainties. The first three digits of the $-\log \mathcal{L}$ values are the same for all runs presented in this table, so we omitted them after the first line for readability. 
\end{table*}

\begin{table*}[!htp]
\begin{center}
\caption{\small Morphological 1$-$100 GeV fit results for M33.}
\begin{tabular}{lccccccc}
\toprule %
Spatial Model  &TS  &TS$_{\rm ext}$  &$-\log \mathcal{L}$ &R.A. (deg)  &Decl. (deg)  &Radius (deg)   & N$_{\rm dof}$ \\
  &  &  & &   &   &    & \\
\hline %
Point source (fixed) &8 &... & 65546.1   &23.462   &30.660   &... &2\\
Point source (free) &23 &... & \hphantom{65}538.5   &$23.62\pm0.05$ &$30.51\pm0.04$   &... &4\\
Disk  &23 & 0.0 & \hphantom{65}538.5 &$23.63\pm0.04$ &$30.51\pm0.04$ &$0.008$  &5 \\
Gaussian &23 & 0.0 & \hphantom{65}538.5 &$23.63\pm0.04$ &$30.51\pm0.04$ &0.003   &5 \\
\textit{Herschel}/PACS map &9 &... & \hphantom{65}545.3 &23.462 &30.662   &... &2 \\
\hline %
\label{spatptlikeM33}
\end{tabular}
\end{center}
Notes: Uncertainties are statistical only. The first two digits of the $-\log \mathcal{L}$ values are the same for all runs presented in this table, so we omitted them after the first line for readability. The epoch for the coordinates is J2000.
\end{table*}

\begin{table*}[!htp]
\begin{center}
\caption{\small Distance, gas masses, gamma-ray luminosity, and average emissivity for M31 and M33.}
\begin{tabular}{lcc}
\toprule 
 Parameter & M31 & M33 \\  
\hline 
$d$ (kpc) & $785\pm25^a$ & $847\pm60^b$ \\
$M_{\rm HI} (10^8 \, \rm M_\odot)$ & $73\pm22^c$ & $19\pm8^d$ \\
$M_{\rm H_2}  (10^8  \, \rm M_\odot)$ & $3.6\pm1.8^e$ & $3.3\pm0.4^d$ \\
$L_{\gamma} (10^{41}  \, \rm ph\,s^{-1})$ & $7.6\pm1.3$ & $< 2.3$ \\
$\overline{q}_\gamma (10^{-25}\, \rm ph\,s^{-1}\,H-atom^{-1})$ & $0.8\pm0.3$ & $<0.9$ \\
\hline
\label{emissM31M33}
\end{tabular}
\end{center}
Notes: (a) \cite{McConnachie05}; (b) \cite{Galleti04}; (c) \cite{Braun09}; (d) \cite{Gratier10}; (e) \cite{Nieten06};  
\end{table*}


\end{document}